\documentclass[aps,prb,superscriptaddress, showkeys, twocolumn]{revtex4-1}
\usepackage{amsmath,amssymb}
\usepackage{graphicx}
\usepackage{dcolumn}
\usepackage{bm}
\usepackage[latin1]{inputenc}
\usepackage{float}
\usepackage{color}
\usepackage{amsmath}
\definecolor{rojo}{rgb}{1,0,0}
\definecolor{verde}{rgb}{0,0.8,0.5}
\definecolor{azul}{rgb}{0,0,1}
\definecolor{rosa}{cmyk}{0,1,0,0}
\usepackage[toc,page]{appendix} 
\usepackage{latexsym}
\usepackage{amsmath}
\usepackage{mathrsfs}
\usepackage{bm}
\usepackage{amsthm}
\usepackage{physics}
\usepackage{adjustbox}
\usepackage{bbm}
\usepackage{amsfonts}
\usepackage{amssymb}
\usepackage{color}
\usepackage{epsfig}
\usepackage{hyperref}
\hypersetup{colorlinks=true, citecolor=blue}
\usepackage{soul}
\usepackage[normalem]{ulem}
\usepackage[cmtip,all]{xy} 
\usepackage{comment}
\usepackage{multirow}
\usepackage{mathtools}
\newcolumntype{L}{>{$}l<{$}}

\newcommand{\longsquiggly}{\xymatrix{{}\ar@{~>}[r]&{}}} 

\begin{document}

\title{{\color{blue}  Response to Comment on: Tunneling in DNA with Spin Orbit coupling}}

\author{Solmar Varela}
\affiliation{Yachay Tech University, School of Chemical Sciences \& Engineering, 100119-Urcuqu\'i, Ecuador}
\affiliation{Yachay Tech University, School of Physical Sciences \& Nanotechnology, 100119-Urcuqu\'i, Ecuador}
\author{Iskra Zambrano}
\affiliation{Yachay Tech University, School of Physical Sciences \& Nanotechnology, 100119-Urcuqu\'i, Ecuador}
\author{Bertrand Berche}
\affiliation{Laboratoire de Physique et Chimie Th\'eoriques, UMR Universit\'e de Lorraine-CNRS 7019 54506 Vandoeuvre les Nancy, France}
\author{Vladimiro Mujica}
\affiliation{School of Molecular Sciences, Arizona State University, Tempe, Arizona 85287-1604, USA}
\author{Ernesto Medina}
\affiliation{Yachay Tech University, School of Physical Sciences \& Nanotechnology, 100119-Urcuqu\'i, Ecuador}
\affiliation{Centro de F\'isica, Instituto 
Venezolano de Investigaciones C\'ientificas (IVIC), Apartado 21827, Caracas 1020 A, Venezuela}%

\date{\today}

\begin{abstract}
The comment in ref.[\onlinecite{WohlmanAharony}] makes a few points related to the validity of our model, especially in the light of the interpretation of Bardarson's theorem: {\it ``in the presence of time reversal symmetry and for half- integral spin the transmission eigenvalues of the two terminal scattering matrix come in (Kramers) degenerate pairs"}. The authors of ref.[\onlinecite{WohlmanAharony}] first propose an ansatz for the wave function in the spin active region and go on to show that the resulting transmission does not show spin dependence, reasoning that spin dependence would violate Bardarson's assertion. Here we clearly show that the ansatz presented assumes spin-momentum independence from the outset and thus just addresses the spinless particle problem. We then find the appropriate eigenfunction contemplating spin-momentum coupling and show that the resulting spectrum obeys Bardarson's theorem. Finally we show that the allowed wavevectors are the ones assumed in the original paper and thus the original conclusions follow. 
We recognize that the Hamiltonian in our paper written in local coordinates on a helix was deceptively simple and offer the expressions of how it should be written to more overtly convey the physics involved. The relation between spin polarization and torque becomes clear, as described in reference ref.[\onlinecite{VarelaZambrano}]. This response is a very important clarification in relation to the implications of Bardarson's theorem concerning the possibility of spin polarization in one dimensional systems in the linear regime.
\end{abstract}

\maketitle
In ref.[\onlinecite{WohlmanAharony}] Aharony et al discuss critical points of our model that allow an opportunity to clarify fine points about notions that have been pointed out in the literature regarding the possibility of spin polarization in one dimensional systems in the linear regime\cite{Bart,BalseiroAharony}. 

Ref. [1] begins to formulate an ansatz for the solution of our Eq.(5) 
\begin{equation}
{\cal H}=\left[\frac{p_x^2}{2m}+V_0\right]{\bf 1}+\alpha\sigma_y p_x,
\label{Hamiltonian}
\end{equation}
by assuming a product wave function in basis of eigenspinors of $\sigma_y$ where $|\Psi_{\mu}(x)=\psi_u(x)|\mu\rangle$. Assuming $\psi_u(x)\propto e^{iQ_{\mu}x}$ they arrive at the energy
\begin{equation}
E=\frac{(Q_{\mu}+k_{so}\mu)^2-k_{so}^2}{2m}+V_0,
\end{equation}
with
\begin{equation}
    Q_{\mu}^{\pm}=-k_{so}\mu \pm q ~{\rm with}~ q=\sqrt{k^2+k_{so}^2-q_0^2},
\end{equation}
where $q_0=2m/\hbar^2$, $k_{so}=m\alpha/\hbar$ and $k^2=2mE/\hbar^2$.
The aforementioned proposal clearly leads to the spinless particle solution, by substituting the proposed $Q_{\mu}$ into the energy, eliminates all dependence on $\mu$. This point makes their transmission computation redundant; they obtain the spinless particle transmission, not surprisingly independent of spin. Furthermore, the solution apparently satisfies Bardarson's theorem since any spin orientation gives the same energy independently of the direction of propagation of the electron.

The latter observation gives a first hint of what has been omitted from this solution which can be drawn from the Hamiltonian above i.e. that there are two sets of Kramers pairs with two different energies (which obey Bardarson's conclusions). Thus, spin and momentum are coupled and the product solution is not forthcoming.

The eigenfunction to Eq.{\ref{Hamiltonian}} is
\begin{equation}
\Psi_{s}=
\left(\begin{array}{c}
     is  \\
  1   
\end{array}\right)e^{i\lambda|q|x},
\label{wavefunction}
\end{equation}
where $s=\pm 1$ which denotes the two possible spin orientations (in $\sigma_z$ basis) and $\lambda=\pm 1$ the two momentum orientations (see ref.\onlinecite{Birkholz}). Substitution of this vector into the eigenvalue equation yields the following energy eigenvalues
\begin{equation}
    E_s^{\lambda}=\frac{\hbar^2 q^2}{2m}-s\lambda\hbar\alpha|q|+V_0,
\end{equation}
that now reflect two Kramers pairs $(s,\lambda)=(+,+)~{\rm and}~(-,-)$ with energy $E_{<}=\hbar^2 q^2/2m-\hbar\alpha|q|+V_0$ and $(s,\lambda)=(+,-)~{\rm and}~(-,+)$ with energy $E_{>}=\hbar^2 q^2/2m+\hbar\alpha|q|+V_0$. The $q$ vectors associated with the eigenfunctions of the Hamiltonian are
\begin{equation}
    q= s k_{so}+ \lambda\sqrt{k^2+k_{so}^2-q_0^2}.
\end{equation}
Note that this is the wavevector of the comment but the form of the wavefunction (Eq.\ref{wavefunction}) involves the quantum number $\lambda$ denoting the wave vector direction. This form is in agreement with our work. Another subtle detail that makes the proposed form of the comment suspect is that the superposition of waves under the barrier does not correspond to equal energies unless the momentum-spin relation is taken into account.

The very subtle difference determines whether this model shows spin polarization under tunneling or not as demonstrated in the comment. For reference work on SO active rings, where these arguments are also applicable see refs.[\onlinecite{Birkholz,Richter,Chatelain,Bolivar}]

Spin polarization in one dimensional systems has been argued to be feasible in the tunneling regime\cite{Bart} in agreement with transport symmetry relations, where there is an energy dependence. In fact if the injection energy is in between $E_{<}$ and $E_{>}$ one expects spin polarization while for the injection energy above $E_{>}$ then both spin orientations are filled and no spin polarization ensues.

The hamiltonian of Eq.(\ref{Hamiltonian}) is deceivingly simple because we wrote in a local coordinate system that rotates because of the constraints imposed by the helix i.e. $k_z=k_y\tan \eta$ (taking $x\rightarrow z$ as in reference [\onlinecite{Varela2016}]) where $\eta$ is the chiral angle of the helix. The system is actually three dimensional with a hamiltonian close to full filling of the form
\begin{eqnarray}
H&=&tR^2\cos^2\eta \left( q_y+q_z\tan\eta\right)^2 \mathbbm{1}_s+\nonumber\\
&&2R\cos\eta~\lambda_{SO}\left( q_y+q_z\tan\eta\right)~s_y,
\end{eqnarray}
applying the relation $q_z=q_y~\tan \eta$ we obtain Eq.(5) of the paper, the Hamiltonian in the paper
\begin{eqnarray}
H&=&tR^2\csc^2\eta~q^2_z~ \mathbbm{1}_s+
2R\csc\eta~ \lambda_{SO}~s_y q_z,
\end{eqnarray}
where again we exchange $x$ in the paper with $z$ in the current notation for consistency ($z$ is along the axis of the helix). This is the Hamiltonian object of the comment of ref.[\onlinecite{WohlmanAharony}], that appears not to involve an orbital degree of freedom for the electron (no orbital angular momentum). To better understand the physics we will eliminate $q_z$ (the vector along the axis of the helix) in favor of $q_y$ using their relationship to obtain (consistent with ref. [\onlinecite{Varela2016}])
\begin{eqnarray}
H=t R^2\sec^{2}{\eta}~q^2_y~\mathbbm{1}_s+2R\sec\eta~\lambda_{SO}~s_y q_y,
\end{eqnarray}
still in local, rotating, coordinates. Further rewriting the Hamiltonian in cylindrical coordinates, where we can identify orbital and spin angular momentum, we get
\begin{equation}
    H=-\beta~\mathbbm{1}_s\partial_{\varphi}^2-i\alpha_{\eta}s_{\varphi}\partial_{\varphi},
    \label{HamiltonianAngle}
\end{equation}
where $\varphi$ is the angle around the helix axis, $s_{\varphi}$ and $\beta=t(R\sec\eta/a)^2$ and $\alpha_{\eta}=2 (R/a)\sec\eta~\lambda_{SO}$. This problem is different from the closed ring since it describes the motion on a helix and thus, periodic boundary conditions do not apply.

The Hamiltonian in the previous equation is nevertheless non-hermitian \cite{Morpurgo} and can be made hermitian by symmetrizing the Hamiltonian in Eq.\ref{HamiltonianAngle}. With the latter procedure we just have to change $\sigma_{\varphi}\partial_{\varphi}\rightarrow \sigma_{\varphi}\partial_{\varphi}-(1/2)\sigma_{\rho}$ so that the hermitian Hamiltonian is
\begin{equation}
    H=-\beta~\mathbbm{1}_s\partial_{\varphi}^2-i\alpha_{\eta}(\sigma_{\varphi}\partial_{\varphi}-\frac{1}{2}\sigma_{\rho}).
    \label{HamiltonianHermitian}
\end{equation}
The Hamiltonian in this form has a very revealing interpretation, and makes now an obvious connection to the conclusions of our paper, since the second term is the kinetic Hamiltonian for a graphene ring $\propto{\bm \sigma}\cdot{\bf p}$, except that here $\mathbbm{\sigma}$ describes real spin, not pseudo-spin. This term will exert a torque on the ring since momentum and spin cannot be kept at a fixed angle on the ring. The torque will disappear if the ring is rotating with the electron momentum. This manifests itself as a pseudo-spin angular momentum in graphene\cite{Bolivar} and it describes the rotation of the ring. 

The same physics applies for the helix but for the real spin, on the helix, the term tends to align momentum (which now circulates around the helix) and spin. Since this is not possible without changing angular momentum then there is a torque on the helix. The latter term is precisely what was computed in the paper as $(1/i\hbar)[s_z,H]={\cal T}$ the torque on the molecule. Addressing the problem in three dimensions brings about new features to the problem regarding the adiabatic/non-adiabatic following of the SO effective magnetic field, which does not arise in the simplified Hamiltonian of Eq.(\ref{Hamiltonian}).
 
 In conclusion, we believe the comment in ref.[\onlinecite{WohlmanAharony}] does not capture the correct spin-momentum coupling present in the model of ref.[\onlinecite{VarelaZambrano}], treating only effectively spinless electron tunneling. We hope our presentation has clarified the issues.

\acknowledgements{We acknowledge fruitfull discussions with Alexander Lopez.}






%


\end{document}